\documentstyle[preprint,tighten,aps]{revtex}

\begin{document}
\draft
\preprint{IFUP-TH 62/97}
\title{
The effective potential in three-dimensional O($N$) models.
}
\author{Andrea Pelissetto and Ettore Vicari}
\address{Dipartimento di Fisica dell'Universit\`a 
and I.N.F.N., I-56126 Pisa, Italy}

\date{\today}

\maketitle

\begin{abstract}
We consider the effective potential in three-dimensional 
models with $O(N)$ symmetry. For generic values of $N$,
and in particular for the physically interesting cases $N=0,1,2,3$,
we determine the six-point and eight-point
renormalized coupling constants which parametrize its small-field
expansion. These 
estimates are obtained from the analysis of their $\epsilon$-expansion,
taking into account the exact results in one and zero dimensions,
and, for the Ising model (i.e. $N=1$), the accurate  
high-temperature estimates in two dimensions. 
They are compared with the available results
from other approaches.
We also obtain corresponding
estimates for the two-dimensional O($N$) models.

\medskip


\end{abstract}

\newpage


\section{Introduction}
\label{introduction}

The effective potential is widely used 
in the field-theoretic description of fundamental interactions 
and phase transitions.
In field theory the effective potential  is the generating functional of the
one-particle irreducible correlation functions at zero external momenta.
In statistical physics it represents the 
free-energy density ${\cal F}$ as a function of the order parameter, which,
for spin models, is the magnetization $M$. 
Its global minimum determines the value of the order parameter,
which characterizes the phase of the system.
In the high-temperature or symmetric phase the minimum is unique with $M=0$. 
As the temperature decreases below the critical value,
the effective potential takes a double-well shape: the order parameter
does not vanish anymore and the system is in the low-temperature
or broken phase\footnote{Actually in the broken phase the double-well
shape is not correct because the effective potential must always
be convex.
In this phase it should present a flat region around the origin.
For a discussion see e.g. Refs.~\cite{OR,M-S}.}.
The equation of state is closely related to the effective potential.
It relates the magnetization $M$
(i.e. the order parameter),  the magnetic field $H$ and the reduced
temperature $t\propto T-T_c$. It is simply given by
\begin{equation}
H={\partial {\cal F}\over \partial M}.
\end{equation}

In this paper we study the effective potential
of O($N$) models. We recall that O($N$) models
describe many important critical
phenomena in nature: liquid-vapour transitions in classical
fluids, the helium superfluid transition, the critical properties of 
isotropic ferromagnetic materials and long polymers.
We will focus mainly on the small-renormalized-field expansion
of the effective potential in the symmetric phase.
Its coefficients are directly related to the 
zero-momentum $n$-point renormalized coupling constants $g_n$.
The four-point coupling $g\equiv g_4$ 
plays an important role in the field-theoretic perturbative expansion
at fixed dimension~\cite{Parisig},
which provides accurate estimates of critical indices and universal 
ratios in the symmetric phase.
In this approach any universal quantity is obtained from a series 
in powers of $g$ ($g$-expansion) which is then
resummed and evaluated at 
the fixed-point value of $g$, $g^*$.
Accurate estimates of $g^*$ have been obtained by 
calculating the zero of the Callan-Symanzik $\beta$-function associated
with $g$ (see e.g. Refs.~\cite{ZJbook,Nickel,LeG-ZJ,Nickel91,Antonenko,G-Z}).
These results have been substantially confirmed by computations using 
different approaches, such as $\epsilon$-expansion~\cite{ONgr2},
high-temperature expansion (see, e.g., 
Refs.~\cite{ONgr,Reisz,Z-L-F,B-C-g,ONgr2}),
Monte Carlo simulations (see, e.g., Refs.~\cite{Tsypin,B-K,K-L}),
$1/N$ expansion~\cite{ONgr,ONgr2}.

Recently there has been a lot of interest in the problem of computing 
the higher-order coupling constants $g_6$, $g_8$, $\ldots$, 
for the Ising model 
(see e.g. Refs.~\cite{T-W,Tsypin,B-B,Z-L-F,Sokolov,G-Z,B-C-g,Morris,K-L,L-F}).
Here we will study the issue for generic values of $N$. We will
obtain satisfactory estimates of 
$g_6$ and $g_8$, or, equivalently, of the ratios
$r_{2j}\equiv g_{2j}/g^{j-1}$, from their 
$\epsilon$-expansion~\cite{Wilson} ($\epsilon\equiv 4-d$)
within the $\phi^4$ theory defined by the action
\begin{equation}
S= \int d^dx\left[ {1\over 2}\partial_\mu \phi(x)\partial_\mu \phi(x)
+ {1\over 2}m_0^2\phi^2 + {1\over 4!}g_0(\phi^2)^2\right].
\label{Sphi4}
\end{equation}

The $\epsilon$-expansion of the fixed-point value of
 $r_{2j}$ can be derived from the 
$\epsilon$-expansion of the equation of state,
which is known to $O(\epsilon^3)$ for the Ising model~\cite{W-Z,N-A}, 
i.e. $N=1$, and to $O(\epsilon^2)$ for generic values of $N$~\cite{B-W-W-e2},
thus leading to series of the same length for $r_{2j}$.
Since the available series are short and have large coefficients
increasing with $j$, their straightforward 
analysis does not provide reliable estimates,
but gives only an indication of the order of magnitude.
A considerable improvement 
can be achieved if one uses the results for $d=0,1,2$, whenever 
they are available.
This idea was employed in Ref.~\cite{eexp} to improve 
the estimates of the critical exponents of the Ising 
and self-avoiding walk models, and in Ref.~\cite{greenfunc}
it was used
to the study of the two-point function.
In Ref.~\cite{ONgr2} it was 
generalized and successfully applied to the determination 
of the zero-momentum four-point renormalized coupling. 
In the present case the basic assumption 
is that the zero-momentum $n$-point renormalized couplings $g_{2j}$,
and therefore the ratios $r_{2j}$,
are analytic and quite smooth in the domain
$4>d>0$ (thus $0 < \epsilon < 4$). 
This can be verified in the large-$N$ limit.
One may then perform a polynomial interpolation among 
the values of $d$ where the constants $r_{2j}$ are known ($d=0,1$) or
for which good estimates are available ($d=2$
for $N=1$, obtained from a high-temperature analysis), 
and then analyze the series of the difference.
This procedure leads to more accurate estimates,
which are consistent with those obtained by the direct analysis
of the original $\epsilon$-series, but have a much smaller uncertainty. 
As a by-product of our analysis we also obtain relatively good estimates of  
$r_{6}$ and $r_8$ for two-dimensional O($N$) models. 

For $N\neq 1$, most of the published results concern 
the renormalized four-point 
coupling constant $g$. As far as we know,
estimates of $g_6$ and $g_8$ have only been obtained from approximate
solutions of the renormalization group equation~\cite{T-W},
and from the analysis of high-temperature series~\cite{Reisz}.
The latter results present a large uncertainty. The former are reported
without errors --- which, in any case, are very difficult to estimate ---
and their reliability is unclear. For instance, the estimates of $g$
that are obtained using this method are in disagreement with 
the results of other computations. Therefore our new 
independent estimates of $r_6$ and $r_8$
for $N\neq 1$ represent our main results, and 
provide an important check of the above-mentioned calculations.

For the Ising model we can compare our results with the
estimates obtained using different approaches:
the $g$-expansion at fixed dimension $d=3$~\cite{Sokolov,G-Z}
that apparently provides the most precise results;
a different analysis of the $\epsilon$-expansion based
on the parametric representation of the equation of state~\cite{G-Z};
approximate solutions of the exact 
renormalization group equation~\cite{T-W,Morris};
high-temperature expansions~\cite{B-C-g,Reisz,Z-L-F};
dimensional expansion around $d=0$~\cite{Bender-etal,B-B};
Monte Carlo simulations~\cite{Tsypin,K-L}. 
Our final results are in good agreement with these estimates 
and their precision is comparable with that of 
the analysis of the $g$-expansion.

The paper is organized as follows.
In Sec.~\ref{sec2} we introduce our notation and give 
some general formulae for 
the small-field expansion of the effective potential.
In Sec.~\ref{sec3} we compute the effective potential in 
the large-$N$ limit, the ratios $r_{2j}$ to $O(1/N)$,
and give the exact values of $r_6$ and $r_8$ 
for $d=1,0$.
Furthermore we present
a high-temperature analysis of the two-dimensional Ising model,
which provides accurate estimates of the first few $r_{2j}$.
In Sec.~\ref{sec4} we present
our analysis of the $\epsilon$-expansion of $r_{2j}$. 
In Sec.~\ref{sec5} we compare our results with other approaches.
In App.~\ref{appa} we give some useful formulae relating
the ratios $r_{2j}$ to the connected Green's functions.
In App.~\ref{appb} we report the $\epsilon$-expansion of 
$r_{2j}$ derived from the  $\epsilon$-expansion
of the equation of state.
In App.~\ref{appc}  some details of the calculations in 
one dimension are given.

\section{Small-field expansion of the effective potential}
\label{sec2}

The free energy per unit volume can be expanded  
in powers of the renormalized magnetization $\varphi$
(i.e. 
the expectation value of the renormalized field $\phi_r=\phi/\sqrt{Z}$):
\begin{equation}
{\cal F}(\varphi) = {\Gamma(\varphi)\over V}=
 {\cal F}(0)+{1\over 2} m^2\varphi^2 + {1\over 4!} 
m^{4-d} g \varphi^4 + 
\sum_{j=3} m^{2j+(1-j)d}
{1\over (2j)!} g_{2j} \varphi^{2j}, 
\label{freeeng}
\end{equation}
where $\Gamma(\varphi)$ is the generating functional of
one-particle irreducible correlation functions at zero external momenta, 
i.e. the effective potential of the renormalized theory.
The mass scale $m$ is 
the inverse of the second-moment correlation length,
i.e. $m=\xi^{-1}$ and
\begin{equation}
\xi^2 = {1\over 2d} {\int dx \ x^2 G(x) \over \int dx\ G(x)},
\end{equation}
where the function $G(x)$ is defined by
\begin{equation}
\langle \phi_\alpha(0) \phi_\beta(x) \rangle
= \delta_{\alpha\beta} G(x).
\end{equation}
By rescaling $\varphi$ as 
\begin{equation}
\varphi = {m^{(d-2)/2}\over\sqrt{g}} z
\label{defzeta}
\end{equation} 
in Eq.~(\ref{freeeng}), 
the free energy can be  written as
\begin{equation}
{\cal F}(\varphi)-{\cal F}(0) = {m^d\over g}A(z),
\label{dAZ}
\end{equation}
where
\begin{equation}
A(z) =   {1\over 2} z^2 + {1\over 4!} z^4 + 
\sum_{j=3} {1\over (2j)!} r_{2j} z^{2j},
\label{AZ}
\end{equation}
and
\begin{equation}
r_{2j} = {g_{2j}\over g^{j-1}} .
\label{r2j}
\end{equation}
In App.~\ref{appa} we give some useful formulae to derive
the constants $r_{2j}$ from the connected Green's functions.

One can show that $z\propto t^{-\beta} M$, and
that the equation of state can be written in the form
\begin{equation}
H\propto t^{\beta\delta} {\partial A(z)\over \partial z}.
\label{eqa}
\end{equation}
This relation can be exploited in order to derive
$A(z)$ from the equation of state, which is usually written 
in the form (see e.g. Ref.~\cite{ZJbook}) 
\begin{equation}
H=M^\delta f(x)
\label{eqstate}
\end{equation}
where $x=t M^{-{1/\beta}}$.
The function $A(z)$ is thus given by
\begin{equation}
{\partial A(z)\over \partial z} = 
h_0 z^\delta f\left(x_0 z^{-{1/\beta}}\right),
\label{daz}
\end{equation}
where the normalization constants $h_0$ and $x_0$ are fixed by
the requirement that
\begin{equation}
A(z) = {1\over 2}z^2 + {1\over 4!} z^4 + O(z^6).
\end{equation}
The ratios $r_{2j}$ are obtained by expanding
$A(z)$ in powers of $z$.
Notice that, since the function $f(x)$
in Eq.~(\ref{eqstate}) is regular at $x=0$
and nonzero, Eq.~(\ref{daz}) implies
$A(z)\sim z^{\delta+1}$ for $z\to\infty$.

In the following we will be interested in the rescaled
effective potential
$A(z)$ and we will calculate the fixed-point values of
the first few coefficients $r_{2j}$ of its small-$z$
expansion.

\section{Exact and high-temperature results}
\label{sec3}

In order to get a qualitative idea of the properties of
the effective potential, we consider its large-$N$ limit. 
It is easy to derive the large-$N$ limit of the 
rescaled effective potential $A(z)$
from the corresponding equation of state~\cite{B-W}:
\begin{equation}
H= M^\delta \left(1 + x \right)^{2/(d-2)}\; ,
\end{equation}
where $\delta=(d+2)/(d-2)$, 
$x=t M^{-{1/\beta}}$ and $\beta=1/2$. 
One finds
\begin{equation}
A(z) = {6\over d} \left[
\left( 1 + {d-2\over 12} z^2\right)^{d/(d-2)} -1 \right],
\end{equation}
from which
\begin{equation}
r_{2j} = {(2j)!\over 2^{2j-2}3^{j-1}j(j-1)}\prod_{i=1}^{j-2}
\left(\epsilon - 2{i-1\over i}\right).
\label{largeNr2j}
\end{equation}     
The constants $r_{2j}$ are $(j-2)$th-order polynomials 
in $\epsilon\equiv 4-d$ that have $j-2$ real zeros at 
$\epsilon = 2(i-1)/i$ with $i=1,...,j-2$. Notice that in the 
limit $j\to\infty$ the zeros have $\epsilon = 2$ 
(i.e. $d=2$) as an accumulation point.
It is interesting to note the form of 
the large-$N$ limit of $A(z)$ for integer $d$:
\begin{eqnarray}
&A(z)= {1\over 2} z^2 + {1\over 24} z^4 \qquad\qquad &{\rm for} \quad d=4,\\
&A(z)= {1\over 2} z^2 + {1\over 24} z^4 + {1\over 864}z^6
\qquad\qquad &{\rm for} \quad d=3,\\
&A(z) = 3\left( e^{z^2/6}-1\right)\qquad\qquad &{\rm for} \quad d=2,\\
&A(z) = 6z^2\left(12 - z^2\right)^{-1}\qquad\qquad &{\rm for} \quad d=1.
\end{eqnarray}
Notice that in the large-$N$ limit, for $d=3,4$ (and in general for 
$d = 2n/(n-1)$ with integer $n\geq 2$) 
the effective potential is a polynomial in $z$, or 
equivalently in $\varphi$. In four dimensions only the 
first two terms are present so that the effective potential 
(in the variable $z$)
coincides with the phenomenological expression of Ginzburg and Landau.
Notice however that the rescaling (\ref{defzeta}) is not strictly defined in
four dimensions since $g\to 0$ in the critical limit. Therefore this 
simple expression is not valid in the original variable $\varphi$ and 
indeed logarithms of the magnetization appear in the four-dimensional
effective potential \cite{Brezinetal-1973}.
In three dimensions also the $\varphi^6$ term is present so that
$\delta=5$. As $d\to2$
all terms become relevant.
Of course this simple behaviour is peculiar of the large-$N$ limit. 
For finite values of $N$ all terms are present in the small-field expansion,
and $\delta$ can only be determined after resumming the series.

One can also derive the $O(1/N)$ correction to Eq.~(\ref{largeNr2j})
from the corresponding $O(1/N)$ correction to the equation of state
calculated\footnote{
We mention the presence of a misprint in the final expression
of the $O(1/N)$ equation of state given in
Eq.~(29) of  Ref.~\cite{B-W}: in the third term of the first line
$(2\pi)^{2-\epsilon/2}$ should be replaced by $2 \pi^{2-\epsilon/2}$.}
 in Ref.~\cite{B-W}.
In $d=3$ one obtains
\begin{eqnarray}
r_6 &=& {5\over 6} \left[ 1 + {12.2556\over N} + O\left( {1\over N^2}\right)
\right] , 
\label{rjlnn6}\\
r_8 &=& -{67.3140\over N}  + O\left( {1\over N^2}\right) , 
\label{rjlnn8} \\
r_{10} &=& {1406.83\over N}  + O\left( {1\over N^2}\right) , 
\label{rjlnn10}  
\end{eqnarray}
etc.... 
For large values of $j$ the coefficients of the $O(1/N)$
term in the $1/N$ expansion of $r_{2j}$ behave approximately as 
$\sim (2j)!(-c)^j$,
where $c$ is a constant: $c\simeq 0.40$. 
The large coefficient of the 
$O(1/N)$ correction in $r_{6}$ indicates that the region 
where Eq.~(\ref{rjlnn6}) 
may be a good approximation corresponds to very
large values of $N$, say $N \gtrsim 100$. 
This is expected to be true also for $r_{2j}$ with larger values of $j$.

The constants $r_{2j}$ can be computed exactly in one and zero dimensions.
These results will be useful in our 
analysis of the $\epsilon$-expansion of $r_{2j}$ as explained in the
introduction.

Some details of the calculations for $d=1$ can be
found in App.~\ref{appc}. Here we only give the results
for $r_6$ and $r_8$.
For $N\geq 1$  we have
\begin{eqnarray}
r_6 &=& 5 - {5 N (N-1)^2 (8 N + 7) \over (N+1) (N+4) (4 N - 1)^2} \ , \\
r_8&=& {175\over3} - 
    {35 N (N-1)^2 (256 N^3 + 3037 N^2 + 1705 N - 588) \over 
         3 (N+1) (N+4) (N+6) (4 N - 1)^3}\ .
\label{r2jd1l}
\end{eqnarray}
For $N\leq 1$ instead
\begin{eqnarray}
r_6 &=& 5,\\
r_8&=& {175\over 3}.
\label{r2jd1s}
\end{eqnarray}
For $N=1$ these expressions agree with the results of Ref.~\cite{B-B}. 
For $N=\infty$ they reproduce 
Eq.~(\ref{largeNr2j}). 
For the Ising model we will also need the value of $r_{10}$, 
which has been computed in Ref.~\cite{B-B}: $r_{10}=1225$.

For $d=0$ and $N\geq 1$, 
using the formulae reported in App.~\ref{appa},
it is easy to obtain\footnote{The calculation is easily
done for the $N$-vector model: in this case one has a single field
$\vec{s}$ with $\vec{s}\cdot\vec{s}=1$ and the Gibbs measure
is simply $d\vec{s}\ \delta(\vec{s}\cdot\vec{s} - 1)$.}
\begin{eqnarray}
r_6 &=& {10(N+8)\over 3(N+4)},\\
r_8&=&{70(N^2+14N+120)\over 3(N+4)(N+6)},\\
r_{10}&=& 
         {280 (10752 + 3136 N + 256 N^2  + 30 N^3  + N^4 ) \over
                  (N+4)^2 (N+6) (N+8)} .
\end{eqnarray}
For $N=1$ these results agree with those of Ref. \cite{B-B}.
It is not clear how to determine the value of $r_{2j}$ for $N=0$. 
Unlike the case $d=1$,
we cannot prove that setting $N=0$ in the
formulae obtained for $N\ge 1$ provides the correct answer.

For the two-dimensional Ising model reliable
estimates of the first few $r_{2j}$ can be obtained from the analysis of their
high-temperature expansion on the lattice.
The basic reason is that 
the leading correction to scaling is analytic, since
the subleading exponent $\Delta$ is expected to be larger 
than one~\cite{Wuising,B-F}. 
Therefore the traditional methods of analysis of high-temperature
series should work well.
The series published in Refs.~\cite{K-Y-T,McKenzie}
for the lattice Ising model with nearest-neighbor interactions
allow us to calculate $r_{2j}$ (more precisely, a high-temperature series
whose value at the critical point is $r_{2j}$) to
17th-order on the square lattice 
(for which $\beta_c={\rm ArcTanh}(\sqrt{2}-1)$)
and to 14th-order on the triangular lattice
(for which $\beta_c={\rm ArcTanh}(2-\sqrt{3})$).
In the analysis of these series  we followed 
Ref.~\cite{ONgr2}, using several types of
approximants, 
Pad\`e, Dlog-Pad\`e and first-order integral 
approximants.
Table~\ref{tabr2j} reports the results obtained 
on the square and on the triangular lattice. 
They are consistent with each other.
Assuming universality, as final estimates
for the two-dimensional Ising model we take
\begin{eqnarray}
r_6 &=& 3.678(2), \label{d2isingr6} \\ 
r_8 &=& 26.0(2),  \label{d2isingr8} \\
r_{10} &=& 275(15) .
\label{d2isingr10}
\end{eqnarray}
The error on the estimate of $r_{2j}$ increases with $j$ and thus the 
analysis of the higher-order coefficients does not lead 
to reliable estimates.
We mention that the high-temperature
analysis of Ref.~\cite{Z-L-F} led\footnote{
Actually Ref.~\cite{Z-L-F} gives an estimate of  
$R_0\equiv \chi_4^2/(\chi_2\chi_6)= (10-r_6)^{-1}$.}
 to $r_6=3.679(8)$,
which is perfectly 
consistent with our estimate.

\section{Analysis of the $\epsilon$-expansion}
\label{sec4}

In this Section we will compute $r_{2j}$ for $j=3,4,5$
using the $\epsilon$-expansion.
The series in $\epsilon$ of $r_{6}$, $r_8$ and $r_{10}$ are reported
in App.~\ref{appb}. They were obtained from the
$\epsilon$-expansion of the equation 
of state~\cite{B-W-W-e2,W-Z,N-A}. 
Since the $\epsilon$-expansion is asymptotic,
it requires a resummation 
to get estimates for $d=3$, i.e. $\epsilon=1$,
which is usually performed assuming its Borel summability. 

The main point of our analysis
is the use of the exact values of $r_{2j}$ for
$d=0,1$ we have reported in the previous Section, 
and, for $N=1$, of the precise two-dimensional estimates
which have been obtained from the analysis of 
high-temperature series, see Eqs. 
(\ref{d2isingr6}), (\ref{d2isingr8}),
(\ref{d2isingr10}). Indeed the constants $r_{2j}$ are expected
to be analytic in the domain $4>d>0$.
This can be explicitly verified in the 
large-$N$ limit where the constants $r_{2j}$ are polynomials in 
$\epsilon$, cf. Eq. (\ref{largeNr2j}).
Moreover it
was implicitly assumed in the 
dimensional expansion around $d=0$ done
in Refs.~\cite{Bender-etal,B-B}.

The idea of the method is the following: consider a generic observable and let 
$R(\epsilon)$ be its expansion in $\epsilon$. Moreover suppose that the 
values of $R$ are known for a set of dimensions 
$\epsilon_1$,...,$\epsilon_k$. In this case 
one may use as zeroth order 
approximation the value for $\epsilon=1$ of the polynomial 
interpolation 
through $\epsilon=0$, $\epsilon_1$,...,$\epsilon_k$ and 
then use the series 
in $\epsilon$ to compute the deviations. 
More precisely, let us suppose that exact values  
$R_{\rm ex}(\epsilon_1)$, $\ldots$, $R_{\rm ex}(\epsilon_k)$ are known 
for the set of dimensions $\epsilon_1$, $\ldots$, $\epsilon_k$, 
$k \ge 2$. Then define 
\begin{equation}
Q(\epsilon) = \sum_{i=1}^k \left[
    {R_{\rm ex}(\epsilon_i) \over (\epsilon - \epsilon_i)}
    \prod_{j=1,j\not=i}^k (\epsilon_i - \epsilon_j)^{-1} \right]
\end{equation}
and 
\begin{equation}
S(\epsilon) = 
   {R(\epsilon) \over \prod_{i=1}^k (\epsilon - \epsilon_i)} - 
   Q(\epsilon) ,
\label{ses}
\end{equation}
and finally
\begin{equation}
   R_{\rm imp}(\epsilon) = \left[ Q(\epsilon) + 
      S(\epsilon) \right]
      \prod_{i=1}^k (\epsilon - \epsilon_i)  .
\label{serieconstrained}
\end{equation}
One can easily verify that the expression
\begin{equation}
   \left[ Q(\epsilon) + S(0) \right]
      \prod_{i=1}^k (\epsilon - \epsilon_i)  
\label{interpolation}
\end{equation}
is the $k$-order polynomial interpolation 
through the points $\epsilon=0,\epsilon_1,...,\epsilon_k$.
The resummation procedure is applied to $S(\epsilon)$
and the final estimate is obtained by computing $R_{\rm imp}(\epsilon=1)$.

Since the series of $r_{2j}$ begins with 
a term of order $\epsilon$, we analyze the quantity $r_{2j}/\epsilon$.
Notice that, as a consequence, the interpolation formula 
(\ref{interpolation}) actually uses the value of the derivative of 
$r_{2j}$  in four dimensions.
If the interpolation is a 
good approximation one should find that the series which gives the 
deviations has smaller coefficients than the original one. 
Consequently one expects that also the errors in the resummation are reduced.
We find that, as expected, the 
coefficients
of the corresponding series $S(\epsilon)$ decrease in size with $k$, 
the number of exact values that are used to constrain the series.
This fact was also shown in Ref.~\cite{ONgr2}
for the case of the four-point renormalized coupling.

The large-$N$ results of Eqs.~(\ref{rjlnn6}-\ref{rjlnn10})
provide further support to our constrained analysis. Indeed 
one may consider the simple polynomial interpolation 
(which uses the values of $r_{2j}$ in $d=0,1$ and the value of its 
derivative in $d=4$)
evaluated at $d=3$, $r^{\rm int}_{2j}$,
and compare its large-$N$ expansion with the exact one.
One finds (for $N\geq 1$)
\begin{eqnarray}
r^{\rm int}_6=&&{5(16N^5+402N^4+1734N^3+539N^2-669N+84)\over
6(N+1)(N+4)(N+8)(4N-1)^2} 
\nonumber \\
=&& {5\over 6}\left[ 1 + {101\over 8N} - {87\over N^2}
+O\left({1\over N^3}\right)
\right] 
\label{r6interp}
\end{eqnarray}
and
\begin{equation}
r^{\rm int}_8 =
-{75.65\over N} + {988.4\over N^2}
+O\left( {1\over N^3}\right) .
\label{r8interp}
\end{equation}
Comparing Eq.~(\ref{r6interp})
with Eq.~(\ref{rjlnn6}), one sees that $r^{\rm int}_6$ 
gives the exact result for $N=\infty$. Moreover,
also the $O(1/N)$ correction is 
closely reproduced: indeed the coefficient of the $O(1/N)$ term 
in $r_6^{\rm int}$ is 
$10.52$ to be compared with the exact value $10.21$ of Eq.~(\ref{rjlnn6}). 
Therefore in the large-$N$ limit $r^{\rm int}_6$ provides an 
estimate of $r_6$ with a relative error which behaves as $0.37/N$: 
$r^{\rm int}_6$ becomes increasingly accurate as $N\to\infty$.
The same discussion applies to $r_8$: the coefficient of the $1/N$ term in
Eq. (\ref{r8interp}) is very close to the exact one $-67.31$,
cf. Eq. (\ref{rjlnn8}). Therefore also 
in this case the interpolation $r^{\rm int}_8$ 
provides 
good estimates of $r_8$: for $N\to\infty$ the relative error is
12\%. 

The analysis of the series $S(\epsilon)$,
cf. Eq.~(\ref{ses}), can be performed
by using the method proposed in Ref.~\cite{LeG-ZJ}, which
is based on the knowledge of the large-order behaviour of the series.
It is indeed known  that the $n$-th coefficient of the series 
behave as $\sim (-a)^n \Gamma(n + b_0 + 1)$ for large $n$.
The constant $a$, which characterizes the singularity of the 
Borel transform  does not depend 
on the specific observable; it is given by~\cite{Lipatov,B-L-Z} 
$a = 3/(N + 8)$.
The coefficient $b_0$ depends instead on the series one considers.
Given a quantity $R$ with series 
\begin{equation}
R(\epsilon)=\sum_{k=0} R_k \epsilon^k,
\end{equation}
we have generated new series $R_p(\alpha,b;\epsilon)$ according to
\begin{equation}
R_p(\alpha,b;\epsilon) = \sum_{k=0}^p 
    B_k(\alpha,b) 
  \int^\infty_0 dt\ t^b\
  e^{-t} {u(\epsilon t)^k \over \left[1 - u(\epsilon t) \right]^\alpha}  
\label{RBorel}
\end{equation}
where 
\begin{equation}
   u(x) = { \sqrt{1 + a x} - 1\over \sqrt{1 + a x} + 1}.
\end{equation}
The coefficients $B_k(\alpha,b)$ are determined by the requirement
that the expansion in $\epsilon$ of $R_p(\alpha,b;\epsilon)$
coincides with the original series. For each $\alpha$, $b$ and $p$
an estimate of $R$ is simply given by $R_p(\alpha,b;\epsilon=1)$.

For the Ising model, where the available series
are of order $O(\epsilon^3)$, we followed Ref.~\cite{ONgr2} in order
to derive the estimates and their uncertainty.
We determine an integer value of $b$, $b_{\rm opt}$, such that
\begin{equation} 
R_3(\alpha,b_{\rm opt};\epsilon=1)\approx R_2(\alpha,b_{\rm opt};\epsilon=1)
\end{equation}
for $\alpha < 1$.
$b_{\rm opt}$ is the value of $b$ such that the estimate
from the series to order $O(\epsilon^3)$ is essentially identical 
to the estimate 
from the series to order $O(\epsilon^2)$. In a somewhat arbitrary way 
we have then considered as our final estimate 
the average of $R_p(\alpha,b;\epsilon=1)$ with 
$-1 < \alpha \le 1$ and $-2 + b_{\rm opt} \le b \le 2 + b_{\rm opt}$.
The error we report is the variance of the values of
$R_3(\alpha,b;\epsilon=1)$
with $-1 < \alpha \le 1$ and 
$\lfloor b_{\rm opt}/3 - 1\rfloor \le b
     \le \lceil  4 b_{\rm opt}/3 + 1\rceil$.
This procedure is {\em ad hoc}, but provides estimates that are all 
consistent among each other. In order to test the method, 
in Ref. \cite{ONgr2}, the procedure was
applied to the determination of the critical indices and it provided 
estimates and error bars in substantial agreement with the results
of other authors. Therefore we believe that our error bars are reasonable,
although one should be cautious in giving them the standard statistical
meaning.

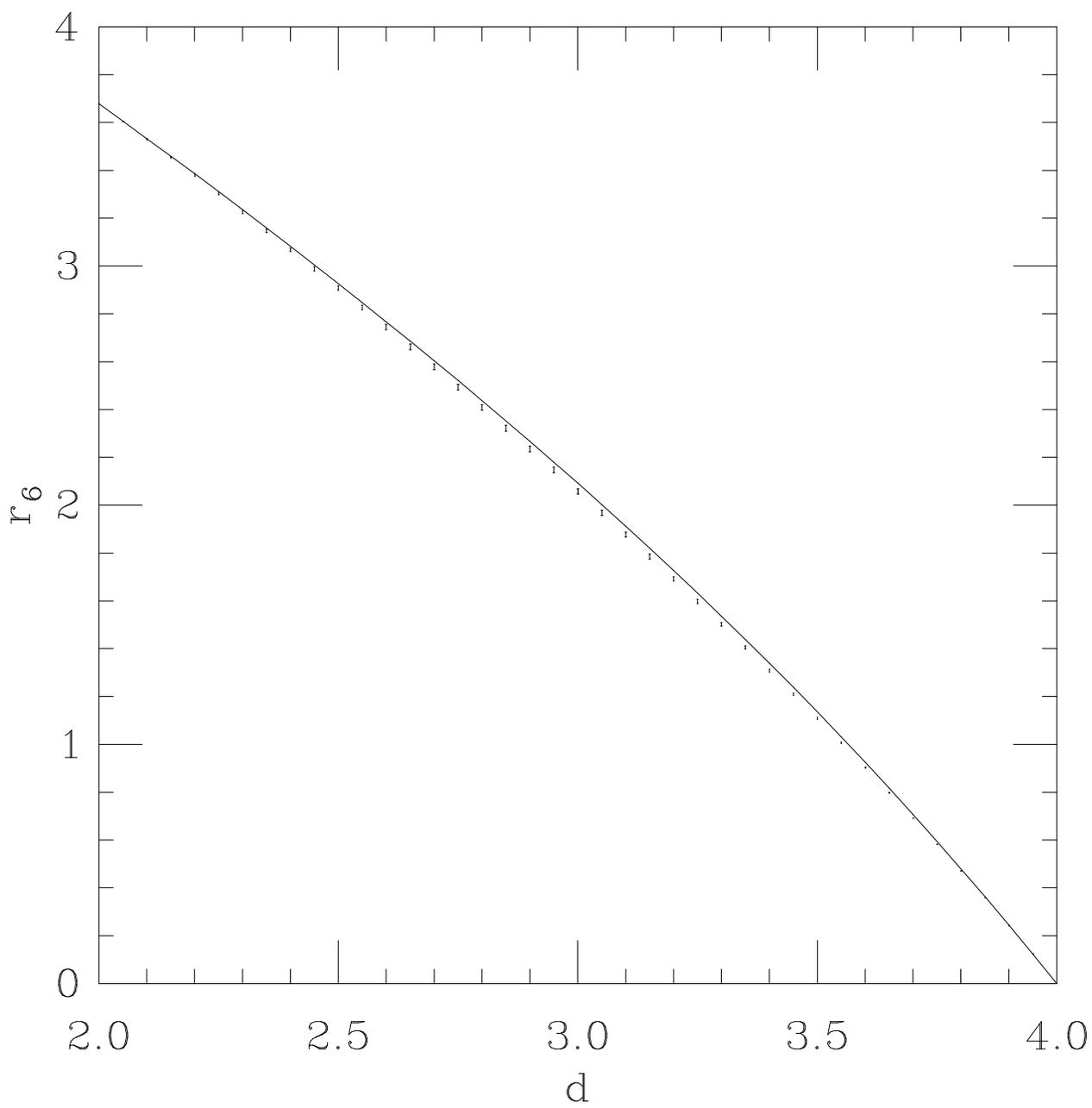
\begin{figure}
\caption{
For the Ising model we plot $r_6$  as a function
of $d$. The continuous line represents the
polynomial interpolation through the known results
at $d=4,2,1,0$. The bars are the results 
(with their uncertainty) of our constrained
($d=2,1,0$) $\epsilon$-expansion analysis. 
}
\label{figr6}
\end{figure}

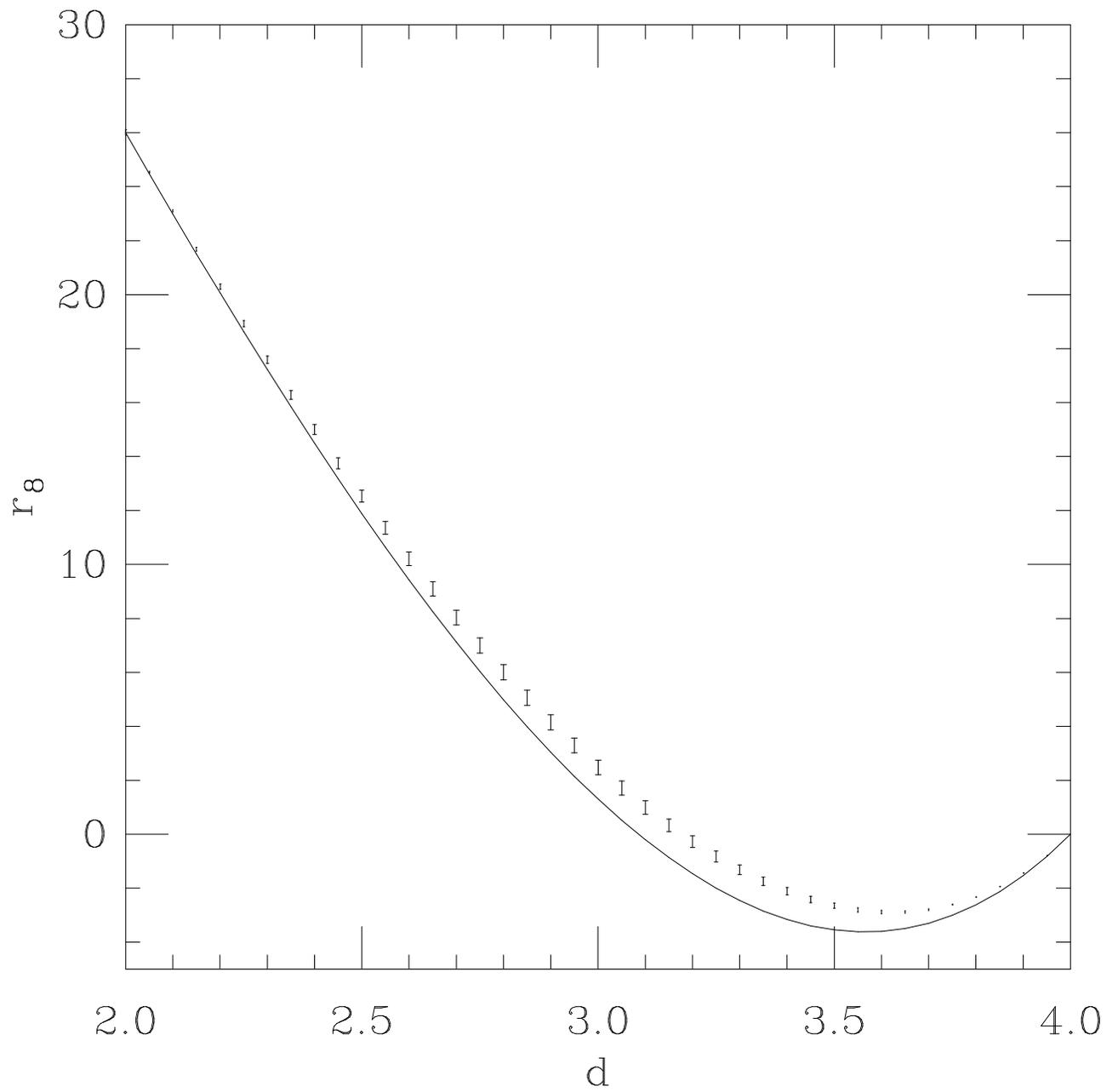
\begin{figure}
\caption{
For the Ising model we plot $r_8$  as a function
of $d$. The continuous line represents the
polynomial interpolation through the known results
at $d=4,2,1,0$. The bars are the results 
(with their uncertainty) of our constrained
($d=2,1,0$) $\epsilon$-expansion analysis. 
}
\label{figr8}
\end{figure}

The results of our analysis for the Ising model, corresponding to $N=1$, 
are presented in Table~\ref{d3N1}. We report 
various estimates of $r_6$, $r_8$ and $r_{10}$ obtained
from an unconstrained analysis and constrained analyses in
various dimensions.
They are all consistent. As expected, the error decreases
when additional lower dimensional values are used to constrain the 
analysis:
the error of the unconstrained analysis is approximately 
an order of magnitude larger than the error of our best result
that uses the known values at $d=0,1,2$.
In Figs.~\ref{figr6} and \ref{figr8} we show respectively
$r_6$ and $r_8$ as a function of $d$. There we plot
the polynomial interpolations through the known values
of $r_6/\epsilon$
at $d=4,2,1,0$, and  the results of our constrained ($d=2,1,0$)
$\epsilon$-expansion analysis. 
Their comparison shows how well the polynomial
interpolation works.

We have also repeated the analysis in two dimensions. 
In this case, of course,
it is more difficult to get precise estimates: the unconstrained 
expansion gives results with large errors and it is therefore
practically
useless. Better estimates are obtained constraining the expansion
in one and zero dimensions. The results for these two cases are reported in 
Table~\ref{tabr2j}.
The final estimates for $r_6$ and $r_8$ are in very good agreement
with the much more precise results obtained from the 
high-temperature analysis. The result for $r_{10}$ is instead 
significantly lower than the high-temperature estimate. 
It should be noted however that the series in $\epsilon$ for $r_{10}$
has very large coefficients and the estimates show large fluctuations 
with the parameters $b$ and $\alpha$. Therefore it is not clear 
if our algorithm to determine the error bars is working properly here.
For this reason, in this case
 we believe the high-temperature estimate to be more reliable
than the $\epsilon$-expansion result.

For generic values of $N$, the series
is one order shorter and we have only two non-trivial terms.
The procedure we presented above cannot be applied and we used a 
different method. For each value of $b$, averaging over 
$-1\le\alpha \le 1$, we 
obtain three estimates: the first one is the result of the 
analysis of the unconstrained series, the second and the third one 
the results from the series constrained respectively in $d=1$ and 
in $d=1,0$. We then compute the weighted average and the 
corresponding $\chi^2$.
The optimal value $b_{\rm opt}$ is chosen as the value of $b$ with the 
smallest $\chi^2$. Once $b_{\rm opt}$ has been chosen, 
we calculated mean values and relative spreads varying $b$ and $\alpha$
according  to the algorithm we presented above. 
Again, we must stress that the procedure is completely arbitrary, and,
since the series has only two terms, one should be very cautious 
in interpreting the spread of the approximants as an error bar. 
Our three-dimensional results for
$r_6$ and $r_8$ are reported in Table~\ref{d3Nvr68}.
There we report the values of
the simple polynomial interpolation
$r^{\rm int}_{2j}$ (see 
Eq.~(\ref{r6interp}) for $r_6^{\rm int}$),
and the results
from an unconstrained analysis and constrained analyses in
various dimensions (the errors reported are the 
spreads of the approximants varying $b$ and $\alpha$).
The corresponding results
for $d=2$ are presented in Table~\ref{d2Nvr68}.

To understand the reliability 
of our procedure we
applied the above analysis to the case $N=1$,
i.e. to its $O(\epsilon^2)$ series of $r_{2j}$ constrained
at $d=0,1$, and compared the results\footnote{
For $N=1$ we have also applied this method using the two-dimensional
results. In this case we consider six estimates, from the analyses of
the series quoted in Table \ref{d3N1}. 
From the series to order
$O(\epsilon^2)$ (resp. $O(\epsilon^3)$) we get 
$r_6 = 2.048(22)$ (resp. $r_6 = 2.061(14)$) and 
$r_8 = 2.9(7)$ (resp. $r_8 = 2.3(5)$).}
(see Table~\ref{d3Nvr68}) with the previous ones reported in Table \ref{d3N1}. 
In three dimensions the estimates of $r_6$ and $r_8$ that have been
obtained using this method are respectively lower (higher) by half 
an error bar (resp. one error bar) than the corresponding estimates 
obtained from the longer series.
So we expect the three-dimensional
results for $r_8$ to be somewhat 
higher than the true result, at least for small values of $N$. 
In two dimensions there is 
good agreement for $r_6$ while $r_8$ differs by one error bar. 
Therefore, for $N=1$, it seems that our procedure gives reasonable 
estimates and error bars.

A rigorous check of the error bars can be done in the large-$N$ limit.
In the case of $r_6$, 
consider the analysis which provides the best estimates, the 
constrained one in $d=1,0$. Here $r_6$ is estimated from the 
expansion 
\begin{equation}
{r_6\over \epsilon}=
{5\over6} \left[ 1 + {10.63 + 0.37 \epsilon\over N} + 
    O(\epsilon^2,1/N^2)\right].
\label{intr6ln}
\end{equation}
This equation shows that inclusion of the $O(\epsilon)$ term gives
an estimate of $r_6$ which is larger than $r_6^{\rm int}$, in contrast 
with what we expect on the basis of the exact result 
(\ref{rjlnn6}). Clearly the deviation $r_6 - r_6^{\rm int}$ is positive
for small values of $\epsilon$ and negative for $\epsilon=1$. Such a behaviour
cannot be reproduced by a single term in $\epsilon$ and thus, at least for 
large values of $N$,  the inclusion of the $O(\epsilon)$ 
in Eq.~(\ref{intr6ln}) worsens the 
final estimate. As a consequence the spread of the approximants largely 
underestimates the correct error.
The previous discussion shows that this is not the case 
for $N=1$ and thus we expect the problem to appear for some intermediate
value of $N$ that is unfortunately unknown.
We have decided to be conservative and we have simply assumed 
that for all values of $N$ the spread of the approximants is an
unreliable estimate of the error. Thus, in Table~\ref{summaryd3Nv}
that summarizes our final results, the error we report for $r_6$
is the difference between the final result and the value of the 
interpolation. This method provides a reasonable estimate of the uncertainty
in the large-$N$ limit, but, for the reasons we mentioned above,
we believe it overestimates the errors
in the small-$N$ region (this is evident in the $N=1$ case).
Of course, the two-dimensional estimates of $r_6$ may have the same problem.
In this case however the spread of the approximants is larger and 
includes $r_6-r_6^{\rm int}$. Therefore, whatever method we use, we get
essentially the same error bars.

The same analysis can be repeated for $r_8$, which is estimated from 
\begin{equation}
{-75.65 + 6.35 \epsilon\over N} + O(\epsilon^2,1/N^2)
\end{equation}
In this case the $O(\epsilon)$ term has the correct sign. For $\epsilon=1$
it provides estimates that differ from the correct result,
Eq. (\ref{rjlnn8}), by 2\%, which is indeed the size of the error bar 
of our final estimates. Therefore the error bars obtained for $r_8$ 
are correct for large values of $N$. Since they are also reasonable for 
$N=1$ we believe them to be reliable for all values of $N$.

\section{Conclusions and comparison with other calculations}
\label{sec5}

We have studied the small-renormalized-field expansion 
of the fixed-point effective potential in
the symmetric phase of three-dimensional O($N$) models. 
The coefficients of this expansion are related to the 
zero-momentum $n$-point renormalized couplings $g_{n}$.
By properly rescaling the effective potential one
may re-express the small-field expansion in terms
of the ratios $r_{2j}\equiv g_{2j}/g^{j-1}$
with $j\geq 3$, cf. Eq.~(\ref{AZ}). 
We have derived the $\epsilon$-expansion of $r_{2j}$ from the
$\epsilon$-expansion of the equation of state which has been computed 
to order $O(\epsilon^3)$ for the Ising model~\cite{W-Z}, and 
to $O(\epsilon^2)$ for $N\neq 1$~\cite{B-W-W-e2}. 
When, as in the case of the ratios $r_{2j}$,
the quantity at hand is expected to be analytic and smooth
as a function of $d$, one may use
exact results (or precise estimates) for lower-dimensional
O$(N)$ models in order to  constrain the analysis of the 
corresponding $\epsilon$-expansion.
For this reason 
we have computed the first few
$r_{2j}$ exactly in one and zero dimensions
for all values of $N$ and we have estimated the same quantities 
in two dimensions for $N=1$ from the high-temperature series
of Refs. \cite{K-Y-T,McKenzie}.
The constrained analyses of the available $\epsilon$-series
of $r_{2j}$,
according to the procedure outlined in Sec.~\ref{sec4},
allowed us to achieve a considerable
improvement with respect to their standard resummation,
and led to satisfactory estimates of the first few $r_{2j}$.
Using the accurate estimates of the fixed-point value
of $g\equiv g_4$ which can be found in the literature
(see Sec.~\ref{introduction}),
one can  extract the fixed-point value of the
zero-momentum $2j$-point renormalized couplings
$g_{2j}$ from the relation $g_{2j}=g^{j-1}r_{2j}$.

Let us compare our results with the available estimates
from other approaches.
For $N\neq 1$  there are not many published results: we are only aware
of the estimates of $g_4$, $g_6$ and $g_8$ presented in Refs.~\cite{T-W,Reisz}.
Table~\ref{summaryd3Nv} presents a summary of all 
the available (as far as we know) estimates of $r_6$ and $r_8$
for several values of $N\neq 1$.
Ref.~\cite{T-W} uses a renormalization-group approach in which
the exact RG equation is approximately solved
(no estimates of the errors are presented there). Ref.~\cite{Reisz} 
instead derives $g_4$ and $g_6$ from their high-temperature 
expansion in the lattice $N$-vector model. 
These estimates are in reasonable agreement with our results.
One can also compare the results of our analysis
for large values of $N$ with
the $1/N$ expansion to $O(1/N)$.
This comparison shows a substantial consistency, although
the region where the $O(1/N)$ approximation of $r_{2j}$
is effective corresponds to very
large values of $N$, say $N \gtrsim 100$.  

We have also computed $r_6$ and $r_8$ in two dimensions. 
We recall that the two-dimensional $N$-vector model is asymptotically
free for $N\geq 3$, and has a Kosterlitz-Thouless
transition for $N=2$.

For $N=1$, the Ising model, many 
works have been devoted to the study of the effective potential,
exploiting various approaches.
Table~\ref{summaryd3N1} presents a summary of all 
the available (as far as we know) estimates\footnote{
When the original reference reports only estimates of $g_{2j}$ 
(see Refs.~\cite{Reisz,B-B,Tsypin,K-L}),
the errors we quote for $r_{2j}$  have been calculated
by considering the estimates of $g_{2j}$ as uncorrelated.} 
of $r_6$, $r_8$ and $r_{10}$.
Apparently the most precise results are those obtained in
Ref.~\cite{G-Z}. They have been derived from the analysis of the $g$-expansion
at fixed dimension 
of the effective potential calculated to 
five loops~\cite{B-B-M-N,H-D}
(in Table~\ref{summaryd3N1} we refer to this approach by
$d=3$ $g$-exp.). 
Ref.~\cite{G-Z} presents also an analysis of the
$O(\epsilon^3)$ expansion using
the parameteric represention of the equation of state
(in Table~\ref{summaryd3N1} we refer to this approach by
$\epsilon$-exp. PREQ).
In this case the $\epsilon$-expansion is used to
estimate the coefficients of the expansion of the function\footnote{
The coefficients of the expansion of $h(\theta)$ 
were estimated by setting $\epsilon=1$ in their $\epsilon$-series.
The reported errors take into account
the uncertainty on the critical exponents 
$\gamma$ and $\beta$, and the error on the normalization
parameter $\rho$ that relates 
$r_{2j}$ and the low-$\theta$ expansion of $h(\theta)$.
Other sources of error are neglected and therefore the final error
may be underestimated. }
$h(\theta)$ characterizing
the parametric representation of the equation of state
(see e.g. Ref.~\cite{ZJbook}).
Then the constants $F_{2j-1}\equiv r_{2j}/(2j-1)!$ are 
obtained using their relation to the expansion of $h(\theta)$.
Precise estimates of $r_{2j}$ have also been obtained 
in Ref.~\cite{Morris} (see also Ref.~\cite{T-W}) by
approximately solving the exact 
renormalization group equation (ERG),
although the estimate of $g$ by the same method is not 
equally good.
Additional results have been obtained from high-temperature 
expansions~\cite{B-C-g,Reisz,Z-L-F}
and Monte Carlo simulations~\cite{Tsypin,K-L} of the lattice Ising model.
The high-temperature results are in substantial 
agreement with the field-theoretic estimates. The apparent small
discrepancy of the estimates of $r_6$ 
of Refs.~\cite{Z-L-F,L-F},
which were obtained by using different lattice formulations
of the Ising model, is probably
due to the presence of confluent singularities which are not properly handled
by standard approximants.
The results of Ref.~\cite{B-C-g} come from an analysis
of $O(\beta^{17})$ series on the cubic lattice, and
they have been obtained by
using the Roskies transform~\cite{Roskies} 
and suitably biased 
integral approximants~\cite{B-C-esponenti} which take into
account the leading confluent singularity.
High-temperature techniques have also been 
used to obtain a dimensional
expansion around $d=0$ ($d$-exp.) of the Green's functions.
The analysis of these series provides estimates of $g_4$ and $g_6$~\cite{B-B}.
The corresponding value of $r_6$ is however smaller than the 
field-theoretic estimates, although the value found for
$g_4$ is in substantial agreement
\footnote{ From the values of $g_4$ and $g_6$ reported for
 the two dimensional model one derives $r_6=3.12(12)$, to be compared
with our strong-coupling result $r_6=3.678(2)$.}. 
The Monte Carlo results do not 
agree with the results of other approaches,
especially those of Ref.~\cite{K-L}.
But one should consider the difficulty
of such calculations on the lattice due to the
subtractions that must be performed to compute the irreducible 
correlation functions.
In Ref.~\cite{Tsypin} estimates of $g_4$ and $g_6$ are obtained
by looking at the probability
distribution of the average magnetization. 
The discrepancy in this case
may come  from the $O(\phi^6)$ 
polynomial approximation of the potential used to fit the data,
or from possible finite-size effects.

Our results for the Ising model are much more precise than those
obtained for $N\not=1$. This is due essentially due to two reasons:
one additional order is known in the $\epsilon$-series of $r_{2j}$;
beside exact results for $d=0,1$, good estimates
have been obtained in $d=2$ by an analysis
of the available high-temperature series,
which can be used in the constrained analysis
of the $\epsilon$-expansion. 
The precision of our results  is
comparable with that obtained from the analysis of the $g$-expansion.
Moreover there is a substantial consistency among
the various approaches whose results are reported
in Table~\ref{summaryd3N1}.

\appendix

\section{}
\label{appa}

In this appendix we give some useful
formulas relating $r_{2j}$ to the connected Green's
functions evaluated at zero momentum.
The coefficients of the expansion of $A(z)$ around $z=0$ can be written
in terms of one-particle irreducible correlation functions
\begin{equation}
\Gamma_{2j}\equiv \Gamma^{(2j)}_{\alpha_1\alpha_1...\alpha_j\alpha_j}
(0,...,0),
\end{equation}
as
\begin{equation}
r_{2j} \equiv {g_{2j}\over g^{j-1}}= {(2j)!\over 2^j3^{j-1}j!}
{(N+2)^{j-2}\over  \prod_{i=2}^{j-1} (N+2i)}
{\Gamma_{2j} \Gamma_2^{j-2}\over \Gamma_4^{j-1}}.
\end{equation}
In terms of the zero-momentum connected Green's functions
\begin{equation}
\chi_{2j} = \sum_{x_2,...,x_{2j}}\langle s_{\alpha_1}(0) s_{\alpha_1}(x_2)
...s_{\alpha_j}(x_{2j-1}) s_{\alpha_j}(x_{2j})\rangle_c,
\end{equation}
one then has
\begin{eqnarray}
r_6 =&& 10 - {5(N+2)\over 3(N+4)}{\chi_6\chi_2\over \chi_4^2},\\
r_8 =&& 280 - {280 (N+2)\over 3(N+4)}{\chi_6\chi_2\over \chi_4^2} 
+{35(N+2)^2\over 9(N+4)(N+6)}{\chi_8\chi_2^2\over \chi_4^3},\\
r_{10} =&& 
15400  
-{7700  (N + 2)\over (N + 4)} {\chi_6 \chi_2\over \chi_4^2}        
+{ 350  (N + 2)^2\over(N + 4)^2} {\chi_6^2 \chi_2^2\over \chi_4^4} 
\nonumber \\ 
 && +{1400 (N + 2)^2\over 3(N + 4) (N + 6)} {\chi_8 \chi_2^2\over \chi_4^3} 
-{35 (N + 2)^3\over 3(N + 4) (N + 6) (N + 8)} 
                   {\chi_{10} \chi_2^3\over \chi_4^4}.
\label{r2jgreen}
\end{eqnarray}

\section{}
\label{appb}

Here we present the $\epsilon$-series of $r_{2j}$ we used
in our analysis.
The $\epsilon$-expansion of $r_{2j}$ can be obtained
from the equation of state which is 
 known to $O(\epsilon^3)$ for the Ising model~\cite{W-Z,N-A,G-Z},
and  to $O(\epsilon^2)$
for generic values of $N$~\cite{B-W-W-e2}.

For generic values of $N$ one finds 
\begin{eqnarray}
r_6&=&{5(N+26)\over 6 (N+8)}\epsilon
- {4.37395N^2+55.6177N+615.008 \over (N+8)^3}\epsilon^2+O(\epsilon^3),\\
r_8&=& -{35(N+80)\over 18(N+8)}\epsilon
+ { 35 ( N^3 + 67.6582 N^2 + 1661.61 N + 11634.7) \over 18(N+8)^3}
\epsilon^2+O(\epsilon^3),\\
r_{10}&=& {35(N+242)\over 3(N+8)}\epsilon
- { 245 (N^3 + 108.389 N^2 + 4780.44 N + 35830.0) \over 12 (N+8)^3}
\epsilon^2+O(\epsilon^3).
\end{eqnarray}

For the Ising model, where one additional order is known,
we have
\begin{eqnarray}
r_6 &=& {5\over 2} \epsilon - {25\over 27} \epsilon^2 
+ \left( {20\zeta(3)\over 9} - {5\lambda\over 9}-{310\over 729}
\right)\epsilon^3 
+O(\epsilon^4),\\
r_8 &=& -{35\over 2} \epsilon + {1925\over 54} \epsilon^2 
- \left({350\zeta(3)\over 9} - {35\lambda\over 3}
+ {18655\over 1458} \right)\epsilon^3 +O(\epsilon^4),\\
r_{10} &=& 315 \epsilon - {13685\over 12} \epsilon^2 
+ \left( 1260\zeta(3) - 420\lambda + {406945\over 324}\right)
\epsilon^3 +O(\epsilon^4),
\end{eqnarray}
where 
\begin{equation}
\lambda=
{1\over 3}\psi'\left({1\over 3}\right) - {2\pi^2\over 9} \approx 
1.17195
\end{equation}
 and $\zeta(3)\approx 1.20206$.

\section{}
\label{appc}

In this appendix we present some details of the computation 
of $r_{6}$ and $r_8$ in $d=1$.
We use the formalism of Ref. \cite{Cucchieri} which is based on the 
expansion of the Boltzmann weight
in hyperspherical harmonics. Explicitly, given a generic 
nearest-neighbour Hamiltonian 
\begin{equation}
H = - \sum_x  V(\vec{s}_x \cdot \vec{s}_{x+1})
\end{equation}
where $\vec{s}_x \cdot \vec{s}_x = 1$, we expand the Boltzmann weight as
\begin{equation}
e^{\beta V(\vec{s}_1 \cdot \vec{s}_2)} = F_N(\beta)
  \left[ 1 + \sum_{k=1}^\infty v_{N,k}(\beta) 
        Y_{N,k}(\vec{s}_1) \cdot Y_{N,k}(\vec{s}_2) \right]
\end{equation}
where $Y_{N,k}(\vec{s})$ are the $O(N)$ hyperspherical harmonics. 
The coefficients $v_{N,k}(\beta)$ depend on $\beta$ and on the explicit form
of the interaction $V(x)$. For the standard nearest-neighbour 
interaction, $V(x) = x$ and 
\begin{equation}
v_{N,k}(\beta) = {I_{N/2+k-1}(\beta) \over I_{N/2-1} (\beta)}
\end{equation}
where $I_n(\beta)$ is a modified Bessel function. 

A rather lengthy computation gives the following results:
\begin{eqnarray}
g_4 &=& 6 - {12 (N-1)\over (N+2)} R_2 ,\\
g_6 &=& 180 + 180 {(N-1)\over (N+2)^2} 
   \left[-5 (N+2) R_2 - {3 N^2\over N+4} R_2^2 R_3 + (7N-4) R_2^2\right] ,
\\
g_8 &=& 12600 + 2520 {(N-1)\over (N+2)^3} \left[
        - 44 (N+2)^2 R_2 
        - 54 {N^2 (N+2)\over N+4} R_2^2 R_3 \right.
\nonumber \\
&& \qquad - 24 {N^3 (N+1) (N+2)\over (N+4)^2 (N+6)} R_2^2 R_3^2 R_4 
          + 6 {N^2 (N+2)\over N+4} R_2^2 R_3^2 + 
            8 (N+2) (17 N - 11) R_2^2 
\nonumber \\
&& \left. \qquad 
         +  12 {N^2 (11 N - 8)\over N + 4} R_2^3 R_3  
         -  18 {N^4 \over (N+4)^2} R_2^3 R_3^2 -
            2 (71 N^2 - 82 N + 20) R_2^3 \right] .
\end{eqnarray}
Here $R_j$ is the critical value of the ratio $m_1/m_j$ where
$m_j$ is the mass in the spin-$j$ channel defined by the large-$x$
behaviour of the spin-$j$ correlation function,
\begin{equation}
  \langle  Y_{N,j}(\vec{s}_0) \cdot Y_{N,j}(\vec{s}_x) \rangle \sim 
    e^{-m_j |x|}
\end{equation}
for $|x|\to\infty$. For the standard nearest-neighbour interaction
we have 
\begin{equation}
R_j =\ \cases{\displaystyle{ {N-1\over j(N + j - 2)} } & 
   \,\, \hbox{\rm for } $N\ge 1$ \cr
   \displaystyle{\vphantom{{N\over j}}} 0 & \,\, \hbox{\rm for } $N \le 1$}
\end{equation}
Substituting in the previous formulae, we get the results
reported in Sec.~\ref{sec3}.


\begin{table}
\caption{
Two-dimensional Ising model.
We report the
estimates of $r_{6}$, $r_8$ and $r_{10}$
obtained
from the analysis of the 17th order strong-coupling
series on the square lattice 
and 14th order series on the triangular lattice (HT),
and from the analyses of the
$\epsilon$-expansion constrained at $d=1$ and at $d=0,1$
($\epsilon$-exp.).
\label{tabr2j}}
\begin{tabular}{ccr@{}lr@{}lcr@{}lr@{}l}
\multicolumn{1}{c}{$$}&
\multicolumn{1}{c}{$$}&
\multicolumn{4}{c}{HT}&
\multicolumn{1}{c}{}&
\multicolumn{4}{c}{$\epsilon$-exp.}\\
\multicolumn{1}{c}{$$}&
\multicolumn{1}{c}{$$}&
\multicolumn{2}{c}{square}&
\multicolumn{2}{c}{triangular}&
\multicolumn{1}{c}{$$}&
\multicolumn{2}{c}{$d=1$}&
\multicolumn{2}{c}{$d=0,1$}\\
\tableline \hline
$r_6$ & $\qquad$ &  3&.677(2) & 3&.678(1) & $\qquad$ &
3&.67(9) &  3&.69(4) \\
$r_8$ & $\qquad$ & 25&.9(5)  & 26&.0(1)  & $\qquad$ & 
24&.2(2.2) &  26&.4(1.0) \\
$r_{10}$ & $\qquad$ & 269&(11) & 279&(11)  & $\qquad$ &
131&(81) &  171&(31) \\
\end{tabular}
\end{table}

\begin{table}
\squeezetable
\caption{Three-dimensional Ising model. 
Estimates of $r_6,r_8$, and $r_{10}$ 
from the polynomial interpolation 
(``int''), 
from an unconstrained analysis
of the $\epsilon$-expansion, ``unc'', and constrained analyses in
various dimensions. For the analyses which use the
estimates in $d=2$
we report two errors: the first one gives the uncertainty of 
the resummation of the series, the second one expresses the 
change in the estimate when the two-dimensional result varies 
within one error bar.
}
\label{d3N1}
\begin{tabular}{cr@{}lr@{}lr@{}lr@{}lr@{}lr@{}lr@{}l}
\multicolumn{1}{c}{}&
\multicolumn{2}{c}{int}&
\multicolumn{2}{c}{unc}&
\multicolumn{2}{c}{$d=1$}&
\multicolumn{2}{c}{$d=0,1$}&
\multicolumn{2}{c}{$d=2$}&
\multicolumn{2}{c}{$d=1,2$}&
\multicolumn{2}{c}{$d=0,1,2$}\\
\tableline \hline
$r_6$ &  2&.092 & 2&.106(79) &  2&.058(35) &  2&.063(24) & 
     2&.059(25+0) & 2&.060(15+1) & 2&.058(11+1) \\ 
$r_8$ &  1&.31 & 0&.4(2.4) &  1&.93(89) &  2&.65(63) & 
     2&.23(60+2) & 2&.53(39+3) & 2&.48(27+5) \\ 
$r_{10}$ & 35& &  $-$98&(120) & $-$7&(67) &  15&(38) &
    $-$7&(41+2) & $-$8&(18+5) & $-$20&(13+7) \\
\end{tabular}
\end{table}

\begin{table}
\squeezetable
\caption{Three-dimensional estimates of $r_6$ and $r_8$
for various
values of $N$ 
from the polynomial interpolation 
(``int''), cf.
Eq.~(\protect\ref{r6interp}) for $r_6$, from
an unconstrained analysis of the
$\epsilon$-expansion and constrained analyses in
various dimensions.
}
\label{d3Nvr68}
\begin{tabular}{ccr@{}lr@{}lr@{}lr@{}lcr@{}lr@{}lr@{}lr@{}l}
\multicolumn{1}{c}{$N$}&
\multicolumn{1}{c}{}&
\multicolumn{8}{c}{$r_6$}&
\multicolumn{1}{c}{}&
\multicolumn{8}{c}{$r_8$}\\
\multicolumn{1}{c}{}&
\multicolumn{1}{c}{}&
\multicolumn{2}{c}{int}&
\multicolumn{2}{c}{unc}&
\multicolumn{2}{c}{$d=1$}&
\multicolumn{2}{c}{$d=0,1$}&
\multicolumn{1}{c}{}&
\multicolumn{2}{c}{int}&
\multicolumn{2}{c}{unc}&
\multicolumn{2}{c}{$d=1$}&
\multicolumn{2}{c}{$d=0,1$}\\
\tableline \hline
0 & $\quad$ &
2&.42 &   2&.08(22)  &  2&.07(12) &  &       & $\quad$ &
$-$7&.8 & 8&(11)      &  6&.4(5.0)  &  &  \\
1 & $\quad$ &
2&.17 &   2&.03(17)  &  2&.01(9)  & 2&.03(6) & $\quad$ &
$-$0&.6 & 5&.2(8.7)   &  4&.8  (3.6) &  4&.7(1.9) \\
2 &  $\quad$ &
2&.05 & 1&.94(14)  &  1&.91(6)  &  1&.94(5) & $\quad$ &
$-$0&.4 & 3&.6(7.0)   &  3&.2  (2.8) &  3&.5(1.3)  \\ 
3 &  $\quad$ &
1&.93 & 1&.86(11)  &  1&.82(5)  &  1&.84(4)  & $\quad$ &
$-$1&.2 & 2&.4(5.8)   &  2&.0  (2.2) &  2&.1(1.0)  \\ 
4 &  $\quad$ &
1&.82 & 1&.76(9)   &  1&.74(4)  &  1&.75(3)  & $\quad$ &
$-$1&.7 & 1&.5(4.9)   &  1&.1  (1.9) &  1&.2(1.0)   \\ 
8 &  $\quad$ &
1&.500&  1&.525(55) &  1&.516(17)&  1&.517(8) & $\quad$ &
$-$2&.5 & $-$0&.7(2.7) &  $-$0&.72(94) & $-$0&.74(44) \\
16 & $\quad$ &
1&.296&  1&.304(26) &  1&.296(6) &  1&.291(1) & $\quad$ &
$-$2&.3 & $-1$&.3(1.4) & $-$1&.29(39) & $-$1&.37(18)  \\
32 & $\quad$ &
1&.104&  1&.115(11) &  1&.112(2) &  1&.108(1) & $\quad$ &
$-$1&.6 & $-1$&.4(7) & $-$1&.18(13) & $-$1&.22(6) \\
48 & $\quad$ &
1&.025&  1&.036(6)  &  1&.032(1) &  1&.029(1) & $\quad$ &
$-$1&.2 & $-1$&.2(5) & $-$0&.96(7) & $-$0&.98(3)  \\
\end{tabular}
\end{table}

\begin{table}
\squeezetable
\caption{Two-dimensional estimates of $r_6$ and $r_8$
for various
values of $N$ 
from the polynomial interpolation
of the known values at $d=4,1,0$,  
from analyses constrained at $d=1$ and at $d=0,1$.
In the large-$N$ limit $r_6=5/3$ and $r_8=35/9$.
}
\label{d2Nvr68}
\begin{tabular}{ccr@{}lr@{}lr@{}lcr@{}lr@{}lr@{}l}
\multicolumn{1}{c}{$N$}&
\multicolumn{1}{c}{}&
\multicolumn{6}{c}{$r_6$}&
\multicolumn{1}{c}{}&
\multicolumn{6}{c}{$r_8$}\\
\multicolumn{1}{c}{}&
\multicolumn{1}{c}{}&
\multicolumn{2}{c}{int}&
\multicolumn{2}{c}{$d=1$}&
\multicolumn{2}{c}{$d=0,1$}&
\multicolumn{1}{c}{}&
\multicolumn{2}{c}{int}&
\multicolumn{2}{c}{$d=1$}&
\multicolumn{2}{c}{$d=0,1$}\\
\tableline \hline
0 &$\quad$& 3&.87 &   3&.7(9) &    &        & $\quad$ &  
10&.5 & 33&(11) &  &  \\ 
1 &$\quad$& 3&.78 &  3&.69(26) &  3&.70(10) & $\quad$ & 
23&.5 & 28&.7(7.5) &  28&.6(2.6) \\ 
2 &$\quad$& 3&.60 & 3&.51(22) &  3&.54(7) & $\quad$ &
20&.9 & 24&.9(6.5) &  25&.1(2.0) \\ 
3 &$\quad$& 3&.38 &  3&.32(19) &  3&.33(6) & $\quad$ &
16&.7 & 20&.3(5.4) &  20&.3(1.7) \\ 
4 &$\quad$& 3&.20 &  3&.15(15) &  3&.15(5) & $\quad$ &
13&.6 & 17&.0(4.5) &  16&.8(1.4) \\ 
8 &$\quad$& 2&.73 &  2&.70(9)  &  2&.71(3) & $\quad$ &
7&.5 &  9&.3(2.2) &   9&.3(7) \\ 
16 &$\quad$& 2&.330 & 2&.34(4)  &  2&.325(5) & $\quad$ &
4&.3 &  5&.1(1.0) &   5&.2(3) \\ 
32 &$\quad$& 2&.045 & 2&.06(2)  &  2&.049(3) & $\quad$ &
3&.4 &  3&.82(40) &  3&.81(12) \\ 
48 &$\quad$& 1&.932 & 1&.95(1)  &  1&.936(2) & $\quad$ & 
3&.3 &  3&.65(22) &  3&.60(7) \\ 
\end{tabular}
\end{table}

\begin{table}
\caption{
Summary of the available
estimates of $r_6$ and $r_8$ for several values
of $N$ in three dimensions.}
\label{summaryd3Nv}
\begin{tabular}{ccr@{}lr@{}lr@{}lr@{}lcr@{}lr@{}lr@{}l}
\multicolumn{1}{c}{N}&
\multicolumn{1}{c}{}&
\multicolumn{8}{c}{$r_6$}&
\multicolumn{1}{c}{}&
\multicolumn{6}{c}{$r_8$}\\
\multicolumn{1}{c}{}&
\multicolumn{1}{c}{}&
\multicolumn{2}{c}{$\epsilon$-exp.}&
\multicolumn{2}{c}{ERG\cite{T-W}}&
\multicolumn{2}{c}{HT\cite{Reisz}}&
\multicolumn{2}{c}{$1/N$}&
\multicolumn{1}{c}{}&
\multicolumn{2}{c}{$\epsilon$-exp.}&
\multicolumn{2}{c}{ERG\cite{T-W}}&
\multicolumn{2}{c}{$1/N$}\\
\tableline \hline
0 & $\qquad$ & 2&.1(3) & & & & & & & $\qquad$ & 
6&(5)& && & \\
2 & $\qquad$ & 1&.94(11) & 1&.83 & 2&.2(6) & && $\qquad$ &  
3&.5(1.3) & 1&.45 & & \\ 
3 & $\qquad$ & 1&.84(9) & 1&.74 & 2&.1(6) & && $\qquad$ &  
2&.1(1.0) & 0&.84 & & \\ 
4 & $\qquad$ & 1&.75(7) & 1&.65 & 1&.9(6) & && $\qquad$ &  
1&.2(1.0) & 0&.33 & & \\ 
8 & $\qquad$ & 1&.52(2) & & & & & 2&.11 & $\qquad$ &  
$-$0&.7(5) & & & $-$8&.41 \\ 
16 & $\qquad$ & 1&.291(5) & & & & & 1&.47 & $\qquad$ &  
$-$1&.4(2) & & & $-$4&.21 \\ 
32 & $\qquad$ & 1&.108(4) & & & & & 1&.15 & $\qquad$ &  
$-$1&.22(6) & & & $-$2&.10 \\ 
48 & $\qquad$ & 1&.029(4) & & & & & 1&.046 & $\qquad$ &  
$-$0&.98(3) & & & $-$1&.40 \\ 
100 & $\qquad$ & 0&.934(3) & 0&.89 & & & 0&.9355 & $\qquad$ &  
$-$0&.575(9) & $-$0&.64 & $-$0&.67 \\ 
\end{tabular}
\end{table}

\begin{table}
\caption{Summary of the available
estimates of $r_6$, $r_8$ and $r_{10}$
for the three-dimensional Ising model. 
The error we quote for the result of our
$\epsilon$-expansion analysis has been calculated
by considering the two errors reported in
the last row of Table~\ref{d3N1} as uncorrelated.} 
\label{summaryd3N1}
\begin{tabular}{cr@{}lr@{}lr@{}l}
\multicolumn{1}{c}{}&
\multicolumn{2}{c}{$r_6$}&
\multicolumn{2}{c}{$r_8$}&
\multicolumn{2}{c}{$r_{10}$}\\
\tableline \hline
$\epsilon$-exp. [this paper] & 2&.058(11) & 2&.48(28) & $-$20&(15)  \\
$\epsilon$-exp. PREQ~\cite{G-Z} & 2&.11(5) & 2&.27(15) & $-$11&.6(7)  \\
$d=3$ $g$-exp.~\cite{G-Z} & 2&.054(7) & 2&.50(25) & $-$22&(15) \\
ERG~\cite{Morris} & 2&.064(36)     & 2&.47(5)   & $-$18&(4) \\
ERG~\cite{T-W}    & 1&.94   & 2&.18   & & \\
HT~\cite{B-C-g} & 1&.99(6) & 2&.7(4) & $-$4&(2) \\
HT~\cite{Z-L-F} & 2&.157(18) & & & &  \\
HT~\cite{L-F}   & 2&.25(9) & & & &  \\
HT~\cite{Reisz} & 2&.5(5) & & & &  \\
$d$-exp~\cite{B-B} & 1&.54(26) & & & & \\
MC~\cite{Tsypin} & 2&.72(23) &  &     & & \\
MC~\cite{K-L} & 3&.26(25)  &12&(2)  & & \\
\end{tabular}
\end{table}

\begin{references}

\bibitem{OR} L.~O'Raifeartaigh, A.~Wipf, and
H.~Yoneyama, Nucl.\ Phys.\ {\bf B271} (1986) 653.

\bibitem{M-S} H.~Mukaida and Y.~Shimada,
Nucl.\ Phys.\ {\bf B479} (1996) 663.

\bibitem{Parisig} G.~Parisi, Carg\`{e}se Lectures (1973), unpublished;
J.\ Stat.\ Phys.\ {\bf 23} (1980) 49.

\bibitem{ZJbook} J.~Zinn-Justin,
``Quantum Field Theory and Critical Phenomena'',
Clarendon Press, Oxford, third edition with corrections, 1997.

\bibitem{Nickel} G.~A.~Baker, Jr.,  B.~G.~Nickel, 
M.~S.~Green and D.~I.~Meiron,
Phys.\ Rev.\ Lett.\ {\bf 36} (1977) 1351;
G.~A.~Baker, Jr., B.~G.~Nickel, and D.~I.~Meiron,
Phys.\ Rev.\ {\bf B17} (1978) 1365.  

\bibitem{LeG-ZJ} J.~C.~Le Guillou, and J.~Zinn-Justin,
Phys.\ Rev.\ Lett.\ {\bf 39} (1977) 95; 
Phys.\ Rev.\ {\bf B21} (1980) 3976.

\bibitem{Nickel91} B.~G.~Nickel, 
Physica {\bf A117} (1991) 189;
D. B. Murray and B.~G.~Nickel, {\em
Revised estimates for critical exponents for the continuum
$n$-vector model in 3 dimensions}, unpublished
Guelph University report (1991).

\bibitem{Antonenko} S.~A.~Antonenko and A.~I.~Sokolov,
Phys.\ Rev.\ {\bf E51} (1995) 1894. 

\bibitem{G-Z} R.~Guida and J.~Zinn-Justin,
Nucl.\ Phys.\ {\bf B489} (1997) 626.

\bibitem{ONgr2} A.~Pelissetto
and E.~Vicari, {\em Four-point renormalized coupling constant and
Callan-Symanzik $\beta$-function in O($N$) models},
cond-mat/9711078, submitted to Nucl. Phys. {\bf B}.

\bibitem{ONgr}M.~Campostrini, A.~Pelissetto,
P.~Rossi and E.~Vicari, Nucl.\ Phys.\ {\bf B459} (1996) 207.

\bibitem{Reisz} T.~Reisz, Phys.\ Lett.\ {\bf 360B} (1995) 77.

\bibitem{Z-L-F} S.~Zinn, S.-N.~Lai, and M.~E.~Fisher,
Phys.\ Rev.\ {\bf E54} (1996) 1176. 

\bibitem{B-C-g} P.~Butera and N.~Comi, 
Phys.\ Rev.\ {\bf E55}  (1997) 6391. 

\bibitem{Tsypin} M.~M.~Tsypin, Phys.\ Rev.\ Lett.\
{\bf 73} (1994) 2015.

\bibitem{B-K} G.~A.~Baker~Jr. and N.~Kawashima,
J.\ Phys.\ {\bf A29} (1996) 7183.

\bibitem{K-L} J.-K.~Kim and D.~P.~Landau, 
Nucl. Phys. B (Proc. Suppl.) {\bf 53} (1997) 706.


\bibitem{T-W} N.~Tetradis and 
C.~Wetterich, Nucl.\ Phys.\ {\bf B422} (1994) 541.

\bibitem{B-B} C.~M.~Bender and S.~Boettcher,
Phys.\ Rev.\ {\bf D48} (1992) 4919;
Phys.\ Rev.\ {\bf D51} (1995) 1875.

\bibitem{Sokolov} A.~I.~Sokolov, V.~A.~Ul'kov and
E.~V.~Orlov, to appear in ``RG96'', and J.\
Phys.\ Studies;
A.~I.~Sokolov, Phys.\ Solid State {\bf 38} (1996) 354.

\bibitem{Morris} T.~Morris,
Nucl.\ Phys.\ {\bf B495} (1997) 477.

\bibitem{L-F} S.-H. Lai and M.~E.~Fisher,
Molec.\ Phys.\ {\bf 88} (1996) 1373.


\bibitem{Wilson} K.~G.~Wilson and M.~E.~Fisher,
Phys.\ Rev.\ Lett.\ {\bf 28} (1972) 240.

\bibitem{W-Z} 
D.~J.~Wallace and R.~P.~K. Zia,
J.\ Phys.\ {\bf C7} (1974) 3480.

\bibitem{N-A} 
J.~F.~Nicoll and P.~C.~Albright,
Phys.\ Rev.\ {\bf B31} (1985) 4576.


\bibitem{B-W-W-e2} E.~Br\'ezin, D.~J.~Wallace, and
K.~G.~Wilson, 
Phys.\ Rev.\ Lett.\ {\bf 29} (1972) 591;
Phys.\ Rev.\ {\bf B7} (1973) 232.

\bibitem{eexp} J.~C.~Le Guillou and J.~Zinn-Justin,
J. Physique {\bf 48} (1987) 19.

\bibitem{greenfunc} M.~Campostrini, A.~Pelissetto,
P.~Rossi and E.~Vicari,
Europhys.\ Lett.\ {\bf 38} (1997) 577; 
cond-mat/9705086, to appear in Phys.\ Rev.\ {\bf E}.


\bibitem{Bender-etal} C.~M.~Bender, S.~Boettcher,
and L.~Lipatov, Phys.\ Rev.\ Lett.\ {\bf 68}
(1992) 3674.

\bibitem{B-W} E.~Br\'ezin and D.~J.~Wallace, 
Phys.\ Rev.\ {\bf B7} (1973) 1967.

\bibitem{Brezinetal-1973}
E. Br\'ezin, J.~C.~Le Guillou and J.~Zinn-Justin,
Phys. Rev. {\bf D8} (1973) 2418.

\bibitem{Wuising}
T.~T.~Wu, B.~M.~McCoy, C.~A.~Tracy, and E.~Barouch,
Phys.\ Rev.\ {\bf B13} (1976) 316.

\bibitem{B-F} M.~Barma, and M.~E.~Fisher,
Phys.\ Rev.\ {\bf B31} (1985) 5954.

\bibitem{K-Y-T} S.~Katsura, N.~Yazaki,
and M.~Takaishi, Can. J. Phys.
{\bf 55}  (1977) 1648.

\bibitem{McKenzie} S.~McKenzie,
Can. J. Phys. {\bf 57} (1979) 1239.

\bibitem{Lipatov} L.~N.~Lipatov, Sov.\ Phys.\ JETP {\bf 72}
(1977) 411.

\bibitem{B-L-Z}
 E.~Br\'ezin, J.~C.~Le Guillou and J. Zinn-Justin, 
Phys. Rev. {\bf D15} (1977) 1544, 1588.

\bibitem{B-B-M-N} C.~Bagnuls,
C.~Bervillier, D.~I.~Meiron and B.~G.~Nickel,
Phys.\ Rev.\ {\bf B35} (1987) 3585.

\bibitem{H-D} F.~J.~Halfkann and V.~Dohm,
Z.\ Phys.\ {\bf B89} (1992) 79.

\bibitem{Roskies} R.~Z.~Roskies,
Phys.\ Rev.\ {\bf B24} (1981) 5305.

\bibitem{B-C-esponenti} P.~Butera and N. Comi,
Phys.\ Rev.\ {\bf B56} (1997) 8212.

\bibitem{Cucchieri} 
A. Cucchieri, T. Mendes, A. Pelissetto and A. D. Sokal,
J. Stat. Phys. {\bf 86} (1997) 581

\end{references}
\end{document}